\documentstyle[aps]{revtex}

\begin{document}
\draft
\preprint{Submitted to Physical Review E Rapid Communications}
\title{Epidemic analysis of the second-order transition in the
Ziff-Gulari-Barshad surface-reaction model}
\author{Christopher A. Voigt and Robert M. Ziff}

\address{Department of Chemical Engineering,
University of Michigan, Ann Arbor, MI 48109-2136}
\date{\today}
\maketitle
\begin{abstract}

We study the dynamic behavior of the Ziff-Gulari-Barshad (ZGB)
irreversible  surface-reaction model around its
kinetic second-order phase transition, using both 
epidemic and poisoning-time analyses. 
We find that the critical point is given by
$p_1 = 0.387\,368\,2 \pm 0.000\,001\,5 $,
which is lower than the previous value.
We also obtain precise 
values of the dynamical critical exponents
$z$, $\delta$, and $\eta$
which provide further numerical 
evidence that this transition is in the same
universality class as directed percolation.

\end{abstract}
\pacs{PACS numbers(s): 05.70.Jk, 11.25.Hf, 64.60.Ak}

There has been a great deal of interest surrounding the critical 
behavior  of non-equilibrium kinetic models,
including directed percolation (DP)
\cite{grass89,GZ}, the  contact process \cite{grassdelatorre}
and various surface-reaction or catalysis models (for review see \cite{albano}).
These models all contain a 
similar continuous (second-order) phase transition from an adsorbing to a vacuum state 
\cite{GLB,grass82}.

 It has been shown that many of these transitions behave in a universal manner even 
though the systems abide by different local rules and are inherently modeling different 
physical systems.
Grassberger \cite{grass89} and Janssen \cite{janssen} postulated that all 
single-component continuous transitions fall into the robust DP
or ``Reggeon field theory" class,
and many numerical simulations have supported this hypothesis 
(i.e., \cite{grass82,JFD,MS}).
Grinstein et al.\ \cite{GLB}
were the first to hypothesize that the specific oxygen-poisoning 
(second-order) transition of the ZGB surface reaction model \cite{ZGB} falls into this class of 
models, and Jensen et al.\ ran simulations which support this conclusion \cite{JFD}.
In this communication, we report on new, very extensive simulations which provide further
support for this hypothesis, and correct an apparent error in the reported value of the
location of that transition.  Assuming
the identification of the ZGB model with the DP class to be exact,
our results give the most accurate values of the DP dynamical critical exponents
to date.

The ZGB model is a simplified model for the irreversible reaction of 
CO (A) and O$_2$ (B$_2$) catalytic 
reaction  on a Pt surface.  The simulation involves the adsorption and reaction of species on 
a square lattice and proceeds via the Langmuir-Hinchelwood mechanism, in which all molecules 
must adsorb before they can react.  The following kinetic scheme is employed,
\begin{eqnarray}
&&	A + * \to  A^*					 \label{eq1}  	     \\
&&	B_2 + * \to 2B^*			 \label{eq2}  \\
&&	A^* + B^* \to AB + 2* 			 \label{eq3} 		 
\end{eqnarray}
where * refers to a lattice site.  A Monte-Carlo algorithm
is  employed where a site is randomly chosen.  If the site is empty, an A will adsorb with 
probability $p$.  With probability $1 - p$, a B$_2$ is adsorbed and instantly dissociates onto that 
site and a randomly chosen neighboring site if the latter is empty.  When a species adsorbs, 
it checks for adjacent neighbors of the opposite species.  If one is present, the two react 
immediately, implying an infinite reaction rate as compared to the adsorption rate.  There are 
two transition points in this system.  At $p_2$, there is a first-order (discontinuous)
transition to  an A-poisoned (saturated) state, and at $p_1$ there is a second-order
(continuous) transition to  a B-poisoned state.  Between these points exists a reactive window
where the system can  reach a steady state and react indefinitely. (For a phase diagram, see
\cite{ZGB}.)  Even within the  window, for finite systems, the system is only metastable as it
can in principle poison by a statistical fluctuation to a non-reactive state.  However, here
the average time to poison $t_p$ grows  exponentially with lattice size $L$, the signature of a
reactive steady state \cite{ARCM}. The value
of  $p_2$ has been accurately determined to be $0.525\,60 \pm 0.000\,01$ \cite{VZ}
using the constant-coverage ensemble algorithm.  This algorithm, however,
is only applicable to finding the location of the
first-order transition.

	Because the second-order transition is a continuous one to a single adsorbing state, 
it is expected to fall into the DP class \cite{grass82}.  Indeed, while the ZGB
model  involves three components (A, B, and vacant sites), at the second-order transition, there are 
rarely A molecules at the surface so it is essentially a two-species model like other members 
of the DP class.  The value of its transition point $p_1$ was first empirically 
observed to be  $0.389 \pm 0.005$ \cite{ZGB}.  A more precise value $0.390\,65 \pm 0.000\,10$
was  obtained by Jensen et al.\ by using an epidemic analysis \cite{JFD}.  However, while
recently  performing some other investigations \cite{VZ}, we found that this value appears to
be somewhat high.  Thus, we carried out new simulations, using the epidemic procedure as
well as a poisoning-time analysis, to re-examine the value of $p_1$ and the related dynamic
critical exponents. 

	The epidemic method was initially used to study the contact process \cite{grassdelatorre} and has 
been successfully applied to determine the critical exponents and the critical point for 
DP \cite{grass89}.  To run the 
epidemic analysis, we started with a large ($1024\times1024$) system completely saturated with B 
except for a single vacant site in the center.  A large system is necessary so that the cluster 
never hits the boundary.  The simulation was run at a set value of $p$ and a reactive cluster 
was grown and watched until the system reverted to a non-reactive adsorbate state, or a 
maximum cutoff time was reached.
  
	The vacant sites (numbering $n_v$) were kept on a list which was randomly accessed 
for each adsorption trial, incrementing the time $t$ by $1/n_v$.  As each cluster grew, the 
quantities of interest were recorded in log$_2$ bins of time.   Since only approximately 3\% of 
all clusters grown reached the last bin, it was necessary to make numerous runs order to 
obtain satisfactory statistics.   For the values $p = 0.387\,36$ and 0.387\,37,
$8\times10^7$ clusters ($N$) 
were grown up to $2^{13}=8192$ time-steps, requiring a total of 200 days of computational time
on a  HP 9000 series UNIX platform.   In the work of \cite{JFD} in contrast, only
$100\,000$ to $250\,000$ clusters were grown up to 1000 timesteps.  Although we could pinpoint
$p_1$ to four significant figures in just a few hours, we decided to carry out extensive runs
in order to find $p_1$ to  six significant digits and to determine the dynamical critical
exponents precisely.

	We measured the three quantities introduced by Grassberger and de la Torre
\cite{grassdelatorre}:   the survival probability $P(t)$, the mean number of vacancies
(averaged over $N$) $n(t)$, and the  mean-square radius of gyration of vacant sites (averaged
over $N$ alive at $t$), $R^2(t)$.  At the  critical point, these are hypothesized to follow
the asymptotic power laws,
\begin{eqnarray}
	P(t)   && \sim t^{-\delta}  \label{eq4}\\	
 n(t)   && \sim t^\eta \label{eq5}\\
 R^2(t) && \sim t^z \label{eq6}
\end{eqnarray}
These exponents follow the hyperscaling relation \cite{grassdelatorre}
\begin{equation}
				dz=2\eta + 4\delta	
\label{eq7}
\end{equation}
where $d$ is the spatial dimension.
 These relationships provide a 
powerful method to determine $p_1$ by evaluating the effects of slightly non-critical values of 
$p$, in which case the resulting behavior deviates from (\ref{eq4}) -- (\ref{eq6}) for
large
$t$.  An  example of this is shown in Fig.~1 where $n(t)$ is plotted for $p = 0.390\,65$,
0.387\,368\,2, and 0.384\,07.  The upper curve is for the value of $p_1$ reported in \cite{JFD},
while the central curve is for the value we find below.  

In order to find the exponents accurately, we consider the local slopes 
which are defined as
\begin{equation}
-\delta(t) = {\ln[P(t)/P(t/2)]/\ln 2 }
\label{eq7prime}
\end{equation}
and similarly for $\eta(t)$ and $z(t)$.  
(Here we used a factor of 2 rather than 5 or 8 of previous work \cite{GZ,JFD},
which we could do because of our higher statistics.)
These are all graphed in Fig.~2 for
$p = 0.387\,36$ and  $0.387\,37$.  The local slopes can be expanded as \cite{grassdelatorre}
\begin{equation}
	\delta(t) = \delta + {a \over t} + {b \over t^{\delta'}}.
\label{eq8}
\end{equation}
If the non-analytic corrections were negligible, then it would be easy to extrapolate the 
critical exponents as a function of $1/t$ as discussed by Grassberger.  However, these corrections 
are rather large and therefore hinder a direct linear extrapolation.  In order to overcome this 
problem, we grew over $5\cdot10^8$ clusters to $2^9$ time-steps so that we could better follow 
the non-analytical trajectory of each curve.  Extrapolating these results to $t \to \infty$, we find
\begin{equation}
\delta = 0.4505 \pm 0.001 , \eta = 0.2295 \pm 0.001 , z = 1.1325 \pm 0.001    . \label{eq9}
\end{equation}
consistent with $\delta = 0.452 \pm 0.008, \eta = 0.224 \pm 0.10$ and $z = 1.133 \pm 0.002$
found in \cite{JFD}.   

For comparison, the updated values recently found by Grassberger and Zhang \cite{GZ} for DP are 
\begin{equation}
	        \delta = 0.451 \pm 0.003 , \eta = 0.229 \pm 0.003 , z = 1.133 \pm 0.002    .            
\label{eq10}
\end{equation}
The precise agreement between (\ref{eq9}) and (\ref{eq10}) leaves little doubt that the ZGB
model is included in the DP class as predicted  by Grinstein et al.\ \cite{GLB}.

	For $p$ away from $p_1$, $n(t)$ follows the scaling behavior \cite{grassdelatorre}		
\begin{equation}
n(t) \sim t^{-\eta} \phi[(p-p_1) t^{1/\nu_{\parallel}}		]
\label{eq11}
\end{equation}     
and similarly for $P(t)$ and $R^2(t)$.  It follows from this equation that
\begin{equation}
{d \ln n \over d \ln p }\Bigg |_{p=p_1}\propto t^{1/\nu_\parallel}
\label{eq12}
\end{equation}    
where $\nu_\parallel$  is the time-domain correlation length exponent.  In Fig.~3, we plot 
the quantity on the left-hand side of the equation above, calculated by  taking
the difference  of $n(t)$ for $p = 0.387\,36$ and $0.387\,37$, vs.\ $\ln t$.
This plot shows that Grassberger's
value for DP $\nu_\parallel  = 1.295 \pm 0.006$ \cite{GZ} is completely
consistent with our data.  

	To determine the precise value of $p_1$, we expand the scaling function $\phi$ as
\begin{equation}
P(t) \approx t^{-\delta}[a + b(p - p_1) t^{1/\nu_\parallel} \ldots ]
\label{eq13}
\end{equation}   
This equation implies that a plot of $P(t)t^\delta$ vs. $t^{1/\nu_\parallel}$   for values $p$ close
to
$p_1$ should yield  straight lines and that a direct linear interpolation of the data from different
values of
$p$ can  be used to find $p_1$ (which corresponds to a horizontal line in this plot) (Fig.~4). 
There is an initial  curvature which is to be expected for small clusters due to finite-cluster
effects.  To  minimize this effect, we have added a constant $c$ to the time that effectively allows
for a higher-order analytic correction term:
\begin{equation}
(t+c)^{-\delta}  \approx t^{-\delta}\left(1 - {\delta c \over t} \right)
\label{eq14}
\end{equation} 
where $c \approx 1.7$ was found to give the best results.  The resulting plot of our data is
shown in Fig.~4.  The statistical fluctuations in each bin are  given by
\begin{equation}
\sqrt{ N_{\rm bin} (  N_{\rm total} - N_{\rm bin} ) \over N_{\rm total} }
\label{eq15}
\end{equation} 
which implies that the largest bins which have the most accurate data also have the greatest 
error (least precision).
Interpolating the two data curves in Fig.~4, we deduce that $p_1$ is given by
\begin{equation}
			    p_1 = 0.387\,368\,2 \pm 0.000\,001\,5.		            	   
\label{eq16}
\end{equation} 
This result is nearly two orders of magnitude more precise than the result of \cite{JFD},
$0.390\,65\pm0.000\,10$, and more than 30 combined error bars lower.  We believe that some error
must have occurred in the simulations or analysis of \cite{JFD}.

	To confirm our value for $p_1$, we also ran a poisoning-time analysis \cite{VZ} of the 
system at its critical point.  Similar methods have been applied to other problems including 
the quantification of finite lattice effects \cite{ARCM,janssen,DM}. 
 To do this, we essentially run
the  opposite dynamic algorithm performed by the epidemic analysis.  We start with a small 
lattice in a fully reactive state (all vacant sites) and set the value of $p$ at our determined
$p_1$.   Periodic boundary conditions are applied and the system is allowed to run until the 
adsorbate B saturates or poisons the system,  causing a global non-reactive state.  When the value
of $p$ is at
$p_1$, it is  expected that the dependence of $t_p$ on $L$ will be power-law, and when $p  \ne p_1$,
the  dependence will be exponential \cite{ARCM,DM}.  We ran this simulation for square lattices of
powers  of 2 in size from $8\times8$ to $64\times64$ for roughly $10^5$ runs each.  Here,
a time-step is  defined as $L^2$ adsorption trials.  Fig. 5 shows the results of our analysis and
it was found  that at $p_1$, the relationship is indeed
\begin{equation}
					t_p \sim L^w					                 
\label{eq17}
\end{equation}
with $w =1.77 \pm 0.02$.  In \cite{VZ},
we observed that $w = 2/z = \nu_\parallel/\nu_\perp$, indicating that the time to  expand
a reactive state scales as the time to contract.
The $z$ implied by this result
is  consistent with the value determined above.  While this method is evidently less 
efficient than the epidemic analysis, it provides a useful confirmation our results for $p_1$ and 
$z$.

	In conclusion, we have provided improved numerical evidence that the ZGB 
oxygen-poisoning transition falls into the larger DP class of nonequilibrium models.
Accepting that that hypothesis is true, which seems certain, our exponents represent the most
accurate values of the DP dynamic critical exponents to date (by a factor of about two).
We also independently confirm that the value of $\nu_\parallel$ for the ZGB model falls into the DP 
class and use it to find a highly accurate and corrected value of $p_1$.

	This material is based upon work supported by the US National Science 
Foundation under Grant No. DMR-9520700.

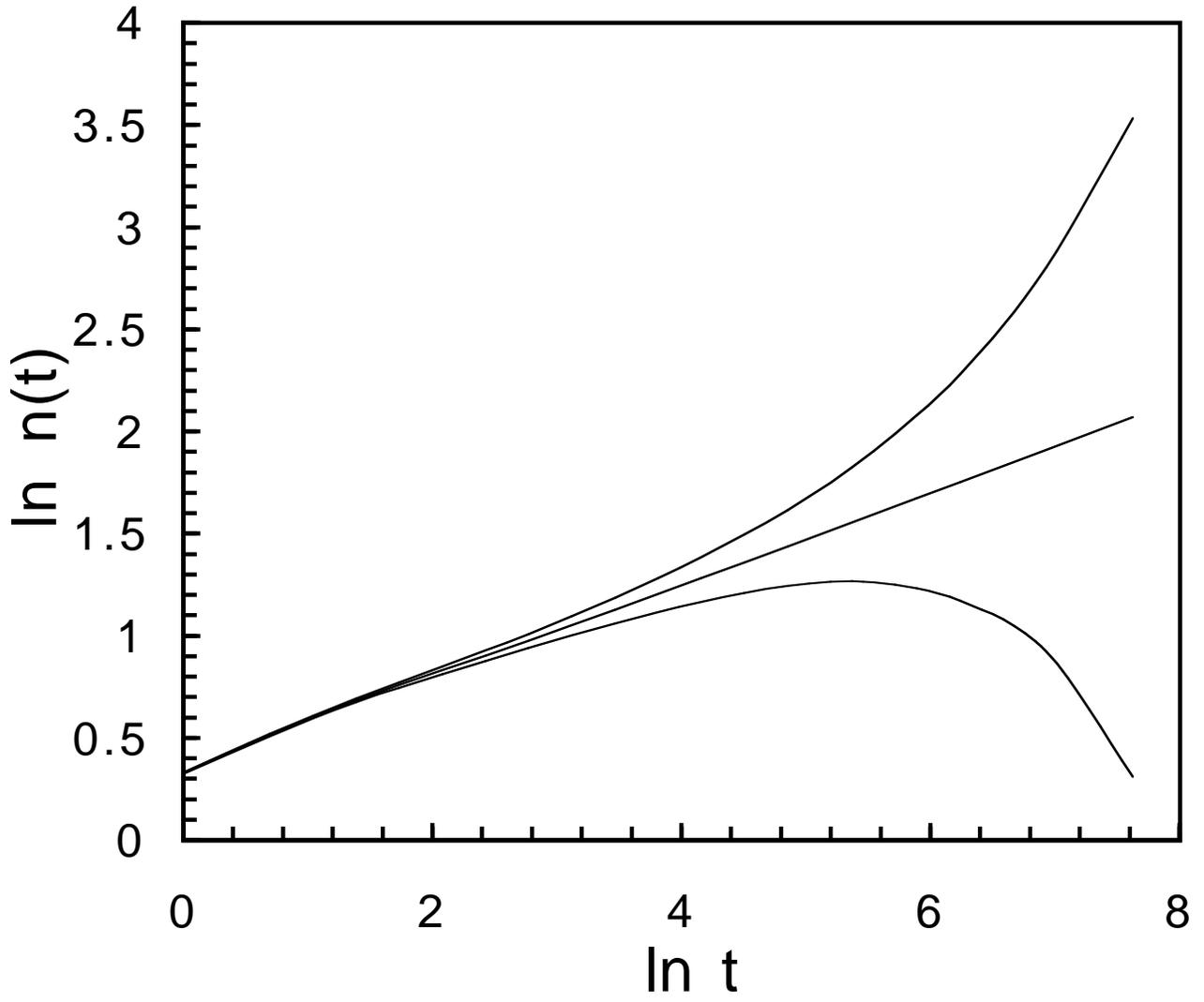
\begin{figure}
\caption{The behavior of the number of vacant sites
$n$ as plotted against time $t$, for $p = 
0.390\,65$, $0.387\,368\,2$, and $0.384\,07$ (top to bottom).}  The upper value is $p_1$ from
\cite{JFD}, and the center is for the value found here.
\end{figure}

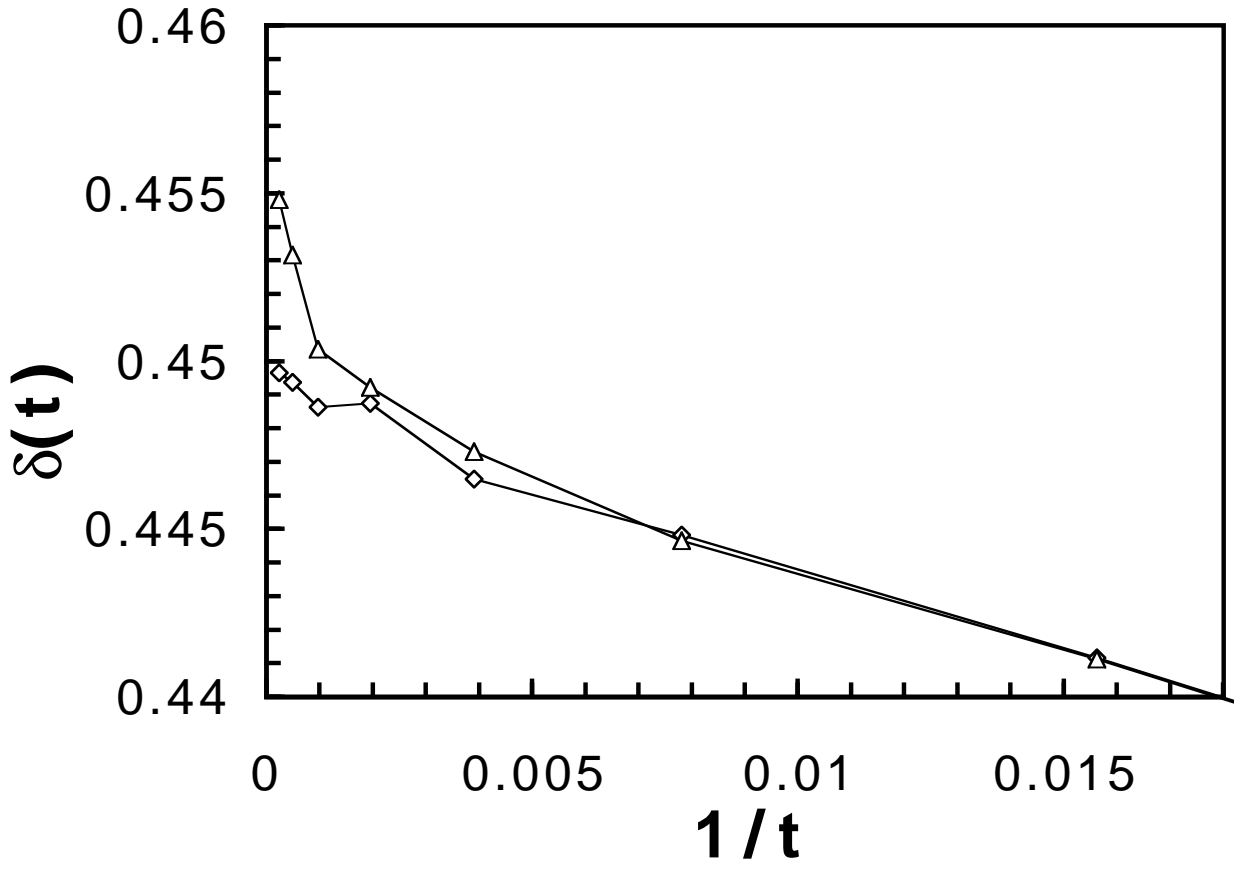
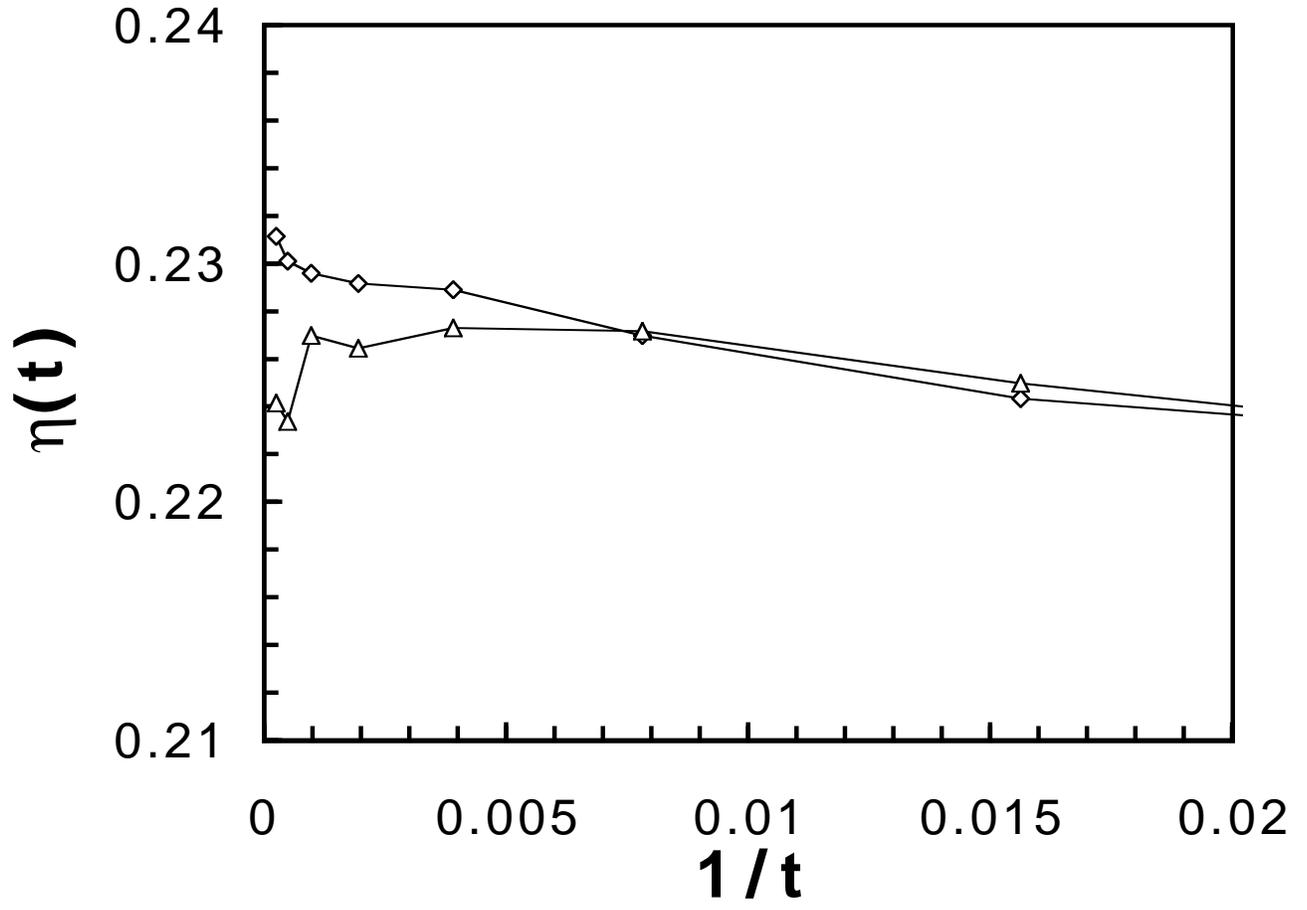
\begin{figure}
\caption{ The three critical exponents derived
from our epidemic analysis: $\delta$ (a),
$\eta$ (b),  and $z$ (c).  These values show super-critical
($\diamond$: $p = 0.387\,37$) and sub-critical
($\triangle$: $p =  0.387\,36$) behavior.
Each of these
lines represents the average of $3.5 \times 10^7$ runs.
The  actual value of $p_1$ falls
between these lines and can be determined by linear interpolation as in Fig.~4.}
\end{figure}

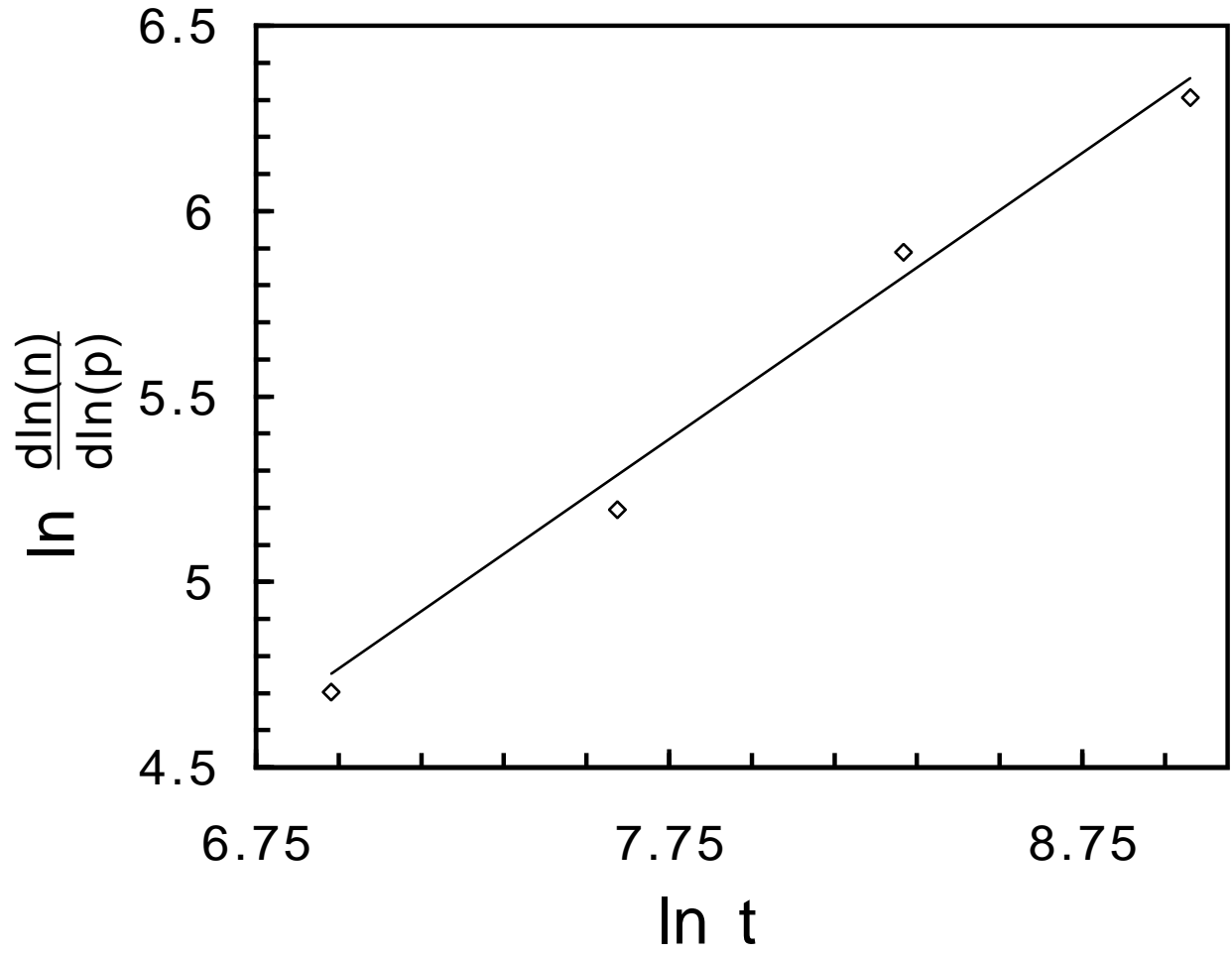
\begin{figure}
\caption{  Plot to determine $\nu_\parallel$ from (13) by using
the values of $p = 0.387\,36$ and $0.387\,37$.  The
line is  for $\nu_\parallel = 1.295$ as determined by
Grassberger for DP.  It can be seen that
the ZGB data is consistent with this value.}
\end{figure}

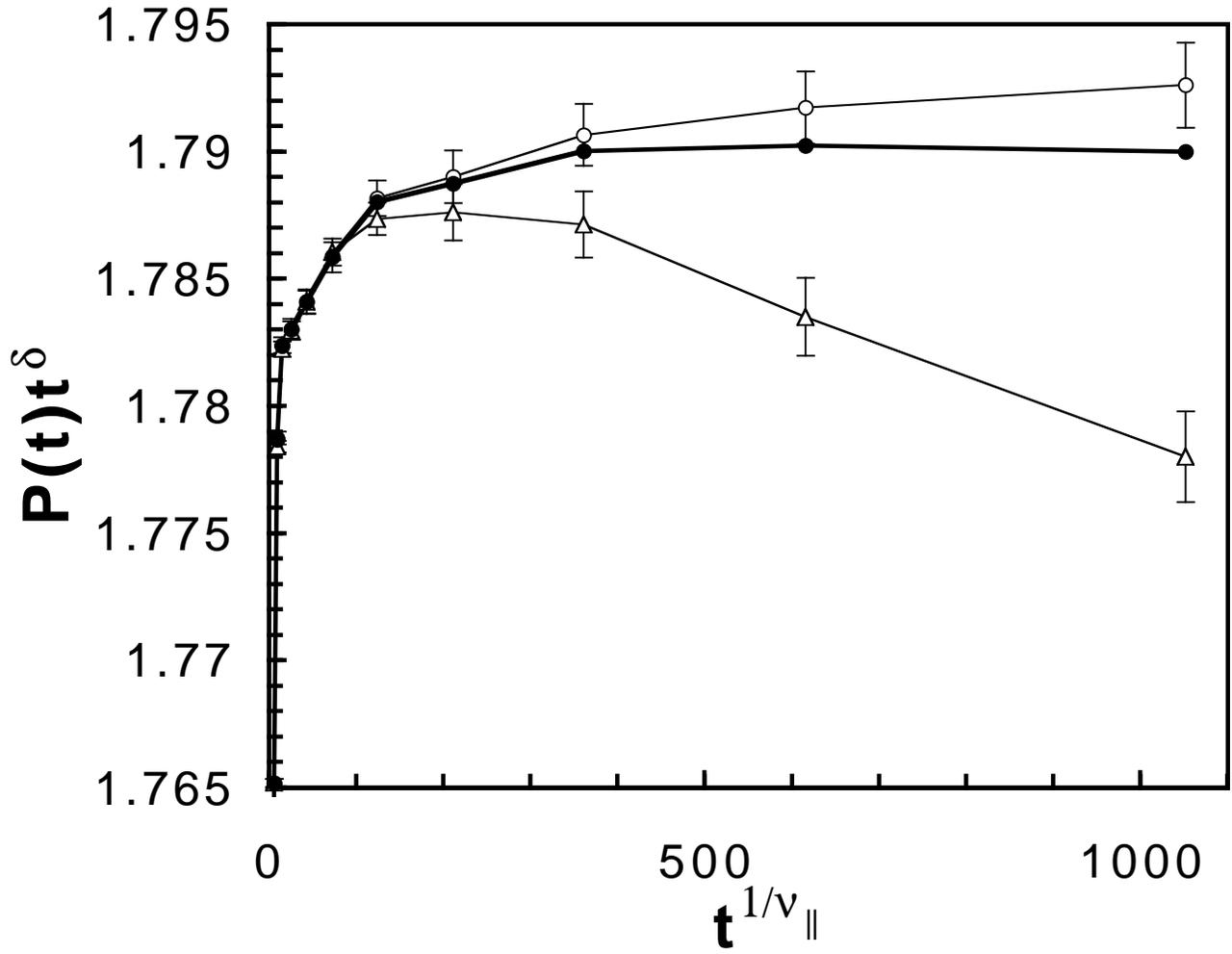
\begin{figure}
\caption{ Plot allowing a linear interpolation for $p_1$
as expressed in equation (14).  The lines 
for $0.387\,36$ ($\triangle$) and $0.387\,37$
($\circ$) represent the sub- and super-critical
behavior  respectively.  The bold line represents
the interpolation for the value, $p_1 =
0.387\,368\,2$.   The error bars were calculated as in (16).
Here, $t$ is offset by an additive
constant $1.7$ to  improve small-time behavior.}
\end{figure}

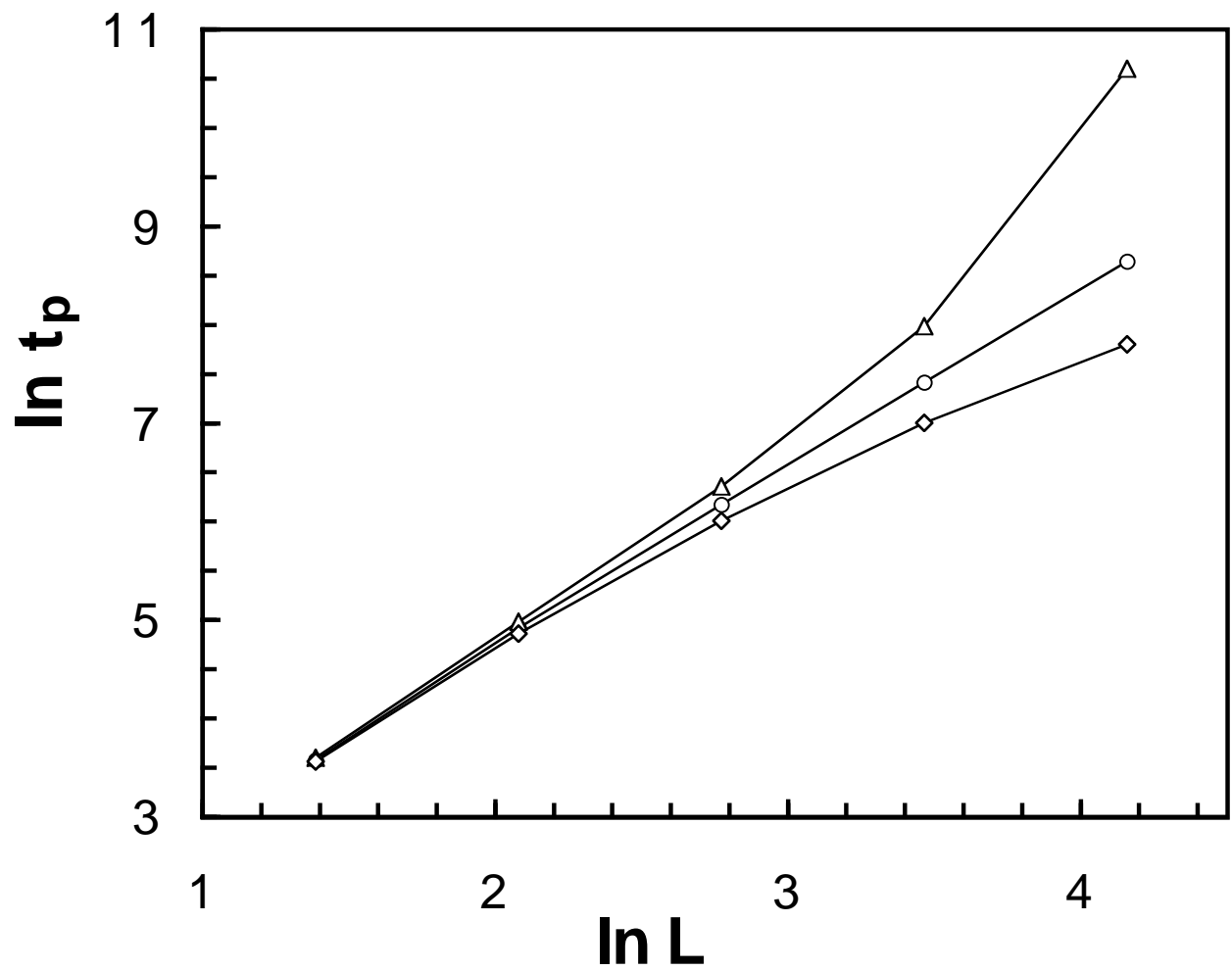
\begin{figure}
\caption{ Results of our poisoning-time analysis for the same
values of $p$ displayed in Fig.~1 ($\triangle$:
$0.390\,65$, $\circ$: $0.387\,368\,2$, $\diamond$: $0.384\,07$).
 This plot demonstrates that the expected power-law  behavior obtains
when our value of $p_1$ is used.}
\end{figure}
\end{document}